\documentclass[sigconf]{acmart}

\usepackage{booktabs} 

\usepackage{balance}

\def\BibTeX{{\rm B\kern-.05em{\sc i\kern-.025em b}\kern-.08emT\kern-.1667em\lower.7ex\hbox{E}\kern-.125emX}}

\usepackage{subcaption}
\usepackage{float}

\copyrightyear{2019} 
\acmYear{2019} 
\setcopyright{acmcopyright}
\acmConference[SIGIR '19]{Proceedings of the 42nd International ACM SIGIR Conference on Research and Development in Information Retrieval}{July 21--25, 2019}{Paris, France}
\acmBooktitle{Proceedings of the 42nd International ACM SIGIR Conference on Research and Development in Information Retrieval (SIGIR '19), July 21--25, 2019, Paris, France}
\acmPrice{15.00}
\acmDOI{10.1145/3331184.3331255}
\acmISBN{978-1-4503-6172-9/19/07}

\title{Unsupervised Neural Generative Semantic Hashing}

\author{Casper Hansen}
\affiliation{
  \city{University of Copenhagen}
}
\email{c.hansen@di.ku.dk}
\author{Christian Hansen}
\affiliation{
  \city{University of Copenhagen}
}
\email{chrh@di.ku.dk}
\author{Jakob Grue Simonsen}
\affiliation{
  \city{University of Copenhagen}
}
\email{simonsen@di.ku.dk}
\author{Stephen Alstrup}
\affiliation{
  \city{University of Copenhagen}
}
\email{s.alstrup@di.ku.dk}
\author{Christina Lioma}
\affiliation{
  \city{University of Copenhagen}
}
\email{c.lioma@di.ku.dk}

\begin{document}

\begin{abstract}
Fast similarity search is a key component in large-scale information retrieval, where semantic hashing has become a popular strategy for representing documents as binary hash codes. Recent advances in this area have been obtained through neural network based models: generative models trained by learning to reconstruct the original documents. We present a novel unsupervised generative semantic hashing approach, \textit{Ranking based Semantic Hashing} (RBSH) that consists of both a variational and a ranking based component. Similarly to variational autoencoders, the variational component is trained to reconstruct the original document conditioned on its generated hash code, and as in prior work, it only considers documents individually. The ranking component solves this limitation by incorporating inter-document similarity into the hash code generation, modelling document ranking through a hinge loss. To circumvent the need for labelled data to compute the hinge loss, we use a weak labeller and thus keep the approach fully unsupervised. 

Extensive experimental evaluation on four publicly available datasets against traditional baselines and recent state-of-the-art methods for semantic hashing shows that RBSH significantly outperforms all other methods across all evaluated hash code lengths. In fact, RBSH hash codes are able to perform similarly to state-of-the-art hash codes while using 2-4x fewer bits. 
\end{abstract}


\maketitle

\section{Introduction}
The task of similarity search consists of querying a potentially massive collection to find the content most similar to a query. In Information Retrieval (IR), fast and precise similarity search is a vital part of large-scale retrieval \cite{wang2018survey}, and has applications in content-based retrieval \cite{lew2006content}, collaborative filtering \cite{koren2008factorization}, and plagiarism detection \cite{henzinger2006finding,stein2007strategies}. Processing large-scale data requires solutions that are both computationally efficient and highly effective, and that work in an unsupervised fashion (because manually labelling massive datasets is unfeasible). Semantic hashing \cite{salakhutdinov2009semantic} is a highly effective class of methods that encode the semantics of a document into a binary vector called a \emph{hash code}, with the property that similar documents have a short Hamming distance between their codes, which is simply the number of differing bits in the codes as efficiently computed by the sum of the XOR operation.
For short hash codes of down to a single byte, this provides a very fast way of performing similarity searches \cite{zhang2010self}, while also reducing the storage requirement compared to full text documents.

Originally, work on semantic hashing focused on generating hash codes for a fixed collection \cite{stein2007principles}, but more modern information needs require querying \emph{unseen} documents for retrieving similar documents in the collection. Modern semantic hashing methods are based on machine learning techniques that, once trained, are able to produce the hash code based solely on the document alone. This can be done using techniques similar to Latent Semantic Indexing \cite{zhang2010laplacian}, spectral clustering \cite{weiss2009spectral}, or two-step approaches of first creating an optimal encoding and then training a classifier to predict this \cite{zhang2010self}. Recent work has focused on deep learning based methods \cite{chaidaroon2017variational, chaidaroon2018deep, nash2018} to create a generative document model. However, none of the methods directly model the end goal of providing an effective similarity search, i.e., being able to accurately rank documents based on their hash codes, but rather just focus solely on generating document representations. 

We present a novel unsupervised generative semantic hashing approach, \textit{Ranking based Semantic Hashing} (RBSH) that combines the ideas of a variational autoencoder, via a so-called variational component, together with a ranking component that aims at directly modelling document similarity through the generated hash codes. The objective of the variational component is to maximize the document likelihood of its generated hash code, which is intractable to compute directly, so a variational lower bound is maximized instead. The variational component is modelled via neural networks and learns to sample the hash code from a Bernoulli distribution, thus allowing end-to-end trainability by avoiding a post-processing step of binarizing the codes. The ranking component aims at learning to rank documents correctly based on their hash codes, and uses weak supervision through an unsupervised document similarity function to obtain pseudo rankings of the original documents, which circumvents the problem of lacking ground truth data in the unsupervised setting. Both components are optimized jointly in a combined neural model, which is designed such that the final model can be used to generate hash codes solely based on a new unseen document, without computing any similarities to documents in the collection. Extensive experimental evaluation on four publicly available datasets against baselines and state-of-the-art methods for semantic hashing, shows that RBSH outperforms all other methods significantly. Similarly to related work \cite{chaidaroon2017variational,chaidaroon2018deep,nash2018}, the evaluation is performed as a similarity search of the most similar documents via the Hamming distance and measured using precision across hash codes of 8-128 bits. In fact, RBSH outperforms other methods to such a degree, that generally RBSH hash codes perform similarly to state-of-the-art hash codes while using 2-4x less bits, which corresponds to an effective storage reduction of a factor 2-4x.

In summary, we \textbf{contribute} a novel generative semantic hashing method, \textit{Ranking based Semantic Hashing} (RBSH), that through weak supervision directly aims to correctly rank generated hash codes, by modelling their relation to weakly labelled similarities between documents in the original space. Experimentally this is shown to significantly outperform all state-of-the-art methods, and most importantly to yield state-of-the-art performance using 2-4x fewer bits than existing methods. 

\section{Related work} \label{s:related-work}
\subsection{Semantic Hashing}
Semantic hashing functions provide a way to transform documents to a low dimensional representation consisting of a sequence of bits. These compact bit vectors are an integral part of fast large-scale similarity search in information retrieval \cite{wang2018survey}, as they allow efficient nearest neighbour look-ups using the Hamming distance. Locality Sensitive Hashing (LSH) \cite{datar2004locality} is a widely known data-independent hashing function with theoretically founded performance guarantees. However, it is general purpose and as such not designed for semantic hashing, hence it empirically performs worse than a broad range of semantic hashing functions \cite{chaidaroon2017variational, chaidaroon2018deep}. 
In comparison to LSH, semantic hashing methods employ machine learning based techniques to learn a \emph{data-dependent} hashing function, which has also been denoted as learning to hash \cite{wang2018survey}.

Spectral Hashing (SpH) \cite{weiss2009spectral} can be viewed as an extension of spectral clustering \cite{ng2002spectral}, and preserves a global similarity structure between documents by creating balanced bit vectors with uncorrelated bits. 
Laplacian co-hashing (LCH) \cite{zhang2010laplacian} can be viewed as a version of binarized Latent Semantic Indexing (LSI) \cite{deerwester1990indexing,salakhutdinov2009semantic} that directly optimizes the Hamming space as opposed to the traditional optimization of Latent Semantic Indexing. Thus, LCH aims at preserving document semantics, just as LSI traditionally does for text representations.
Self-Taught Hashing (STH) \cite{zhang2010self} has the objective of preserving the local similarities between samples found via a $k$-nearest neighbour search. This is done through computing the bit vectors by considering document connectivity, however without learning document features. Thus, the objective of preserving local similarities contrasts the global similarity preservation of SpH. Interestingly, the aim of our RBSH can be considered as the junction of the aims of STH and SpH: the variational component of RBSH enables the learning of local structures, while the ranking component ensures that the hash codes incorporate both local and global structure. 
Variational Deep Semantic Hashing (VDSH) \cite{chaidaroon2017variational} is a generative model that aims to improve upon STH by incorporating document features by preserving the semantics of each document using a neural autoencoder architecture, but without considering the neighbourhood around each document. The final bit vector is created using the median method \cite{weiss2009spectral} for binarization, which means the model is not end-to-end trainable.
Chaidaroon et al. \cite{chaidaroon2018deep} propose a generative model with a similar architecture to VDSH, but in contrast incorporate an average document of the neighbouring documents found via BM25 \cite{robertson1995okapi} which can be seen as a type of weak supervision. The model learns to also reconstruct the average neighbourhood document in addition to the original document, which has similarities with STH in the sense that they both aim to preserve local semantic similarities. In contrast, RBSH directly models document similarities based on a weakly supervised ranking through a hinge loss, thus enabling the optimization of both local and global structure. Chaidaroon et al.\ \cite{chaidaroon2018deep} also propose a model that combines the average neighbourhood documents with the original document when generating the hash code. However this model is very computationally expensive in practice as it requires to find the top-k similar documents online at test time, while not outperforming their original model \cite{chaidaroon2018deep}.
NASH \cite{nash2018} proposed an end-to-end trainable generative semantic hashing model that learns the final bit vector directly, without using a second step of binarizing the vectors once they have been generated. This binarization is discrete and thus not differentiable, so a straight-through estimator \cite{bengio2013estimating} is used when optimizing the model. 

The related work described above has focused on unsupervised text hashing. Direct modelling of the hash code similarities as proposed in this paper has not been explored. For the case of \emph{supervised} image hashing, some existing work has aimed at generating hash codes using ranking strategies from labelled data, e.g., based on linear hash functions \cite{wang2013learning} and convolutional neural networks \cite{zhao2015deep,yao2016deep}. In contrast, our work develops a generative model and utilises weak supervision to circumvent the need for labelled data.

\subsection{Weak Supervision}
Weak supervision has showed strong results in the IR community \cite{dehghani2017neural,nie2018multi,zamani2018sigir,hansen2019neural}, by providing a solution for problems with small amounts of labelled data, but large amounts of unlabelled data. While none of these are applied in a problem domain similar to ours, they all show that increased performance can be achieved by utilizing weak labels. Zamani et al.\ \cite{dehghani2017neural} train a neural network end-to-end for ad-hoc retrieval. They empirically show that a neural model trained on weakly labelled data via BM25 is able to generalize and outperform BM25 itself. A similar approach is proposed by Nie et al.\ \cite{nie2018multi}, who use a multi-level convolutional network architecture, allowing to better differentiate between the abstraction levels needed for different queries and documents. Zamani et al.\ \cite{zamani2018sigir} present a solution for the related problem of query performance prediction, where multiple weak signals of clarity, commitment, and utility achieve state-of-the-art results. 

\section{Ranking based Semantic Hashing}\label{s:ranking-based-semantic-hashing}
We first present an overview of our model, Ranking Based Semantic Hashing (RBSH), and then describe in detail the individual parts of the model. RBSH combines the principles of a variational autoencoder with a ranking component using weak supervision and is an unsupervised generative model. For document $d$, the variational component of RBSH learns a low dimensional binary vector representation $z \in \{0,1\}^m$, called the hash code, where $m$ is the number of bits in the code. RBSH learns an encoder and decoder function, modelled by neural networks, that are able to encode $d$ to $z$ and $z$ to $\hat{d}$, respectively, where $\hat{d}$ is an approximation of the original document $d$. The goal of the encoder-decoder architecture is to reconstruct the original document as well as possible via the hash code. 
Additionally, we incorporate a ranking component which aims to model the similarity between documents, such that the resulting hash codes are better suited for finding nearest neighbours. Specifically, during training RBSH takes document triplets, $(d, d_1, d_2)$, as inputs with estimated pairwise similarities, and through weak supervision attempts to correctly predict either $d_1$ or $d_2$ as being most similar to $d$. Training the model on inputs of various similarities (e.g., from the top 200 most similar documents) enables the model to learn both the local and global structure to be used in the hash code generation.

\begin{figure*}
    \centering
    \includegraphics[width=0.85\linewidth]{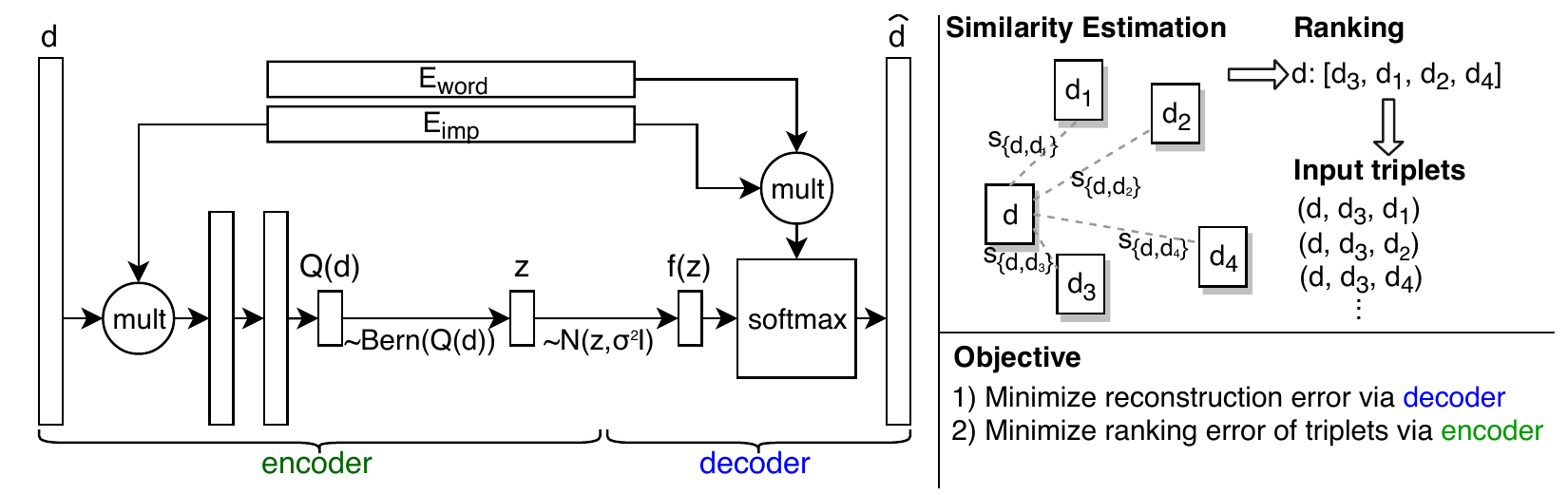}
    \vspace{-10pt}
    \caption{Model overview}
    \label{fig:model}
    \vspace{-10pt}
\end{figure*}

In summary, through the combination of the variational and ranking components the objective of RBSH is to be able to both reconstruct the original document as well as correctly rank the documents based on the produced hash codes. An overview of the model can be seen in Figure \ref{fig:model}.
In the sections below we describe the generative process of the variational component (Section \ref{ss:generative-process-and-objective-function}), followed by the encoder function (Section \ref{ss:encoder-function}), decoder function (Section \ref{ss:decoder-function}), the ranking component (Section \ref{ss:weak-supervision-modelling}), and finally the combined model (Section \ref{ss:combined-components}). 

\subsection{Variational component}
\label{ss:generative-process-and-objective-function}

We assume each document $d$ to be represented as a bag-of-words representation of vocabulary size $V$ such that $d \in \mathbb{R}^V$. We denote the set of unique words in document $d$ as $\mathcal{W}_d$.
For each document we sample a binary semantic vector $z \sim p(z)$ where $p(z_i) = p_i^{z_i}(1-p_i)^{1-z_i}$, which allows the hash codes to be end-to-end trainable, similarly to Shen et al. \cite{nash2018}. For each bit, $p_i$ corresponds to the probability of sampling a 1 at position $i$ and $(1-p_i)$ is the probability of sampling a 0. Thus, $z$ is obtained by repeating a Bernoulli trial $m$ times.
Using the sampled semantic vector, we consider each word as $w_i \sim p(w_i|f(z))$ and define the document likelihood as follows:
\begin{align}\label{eq:doc_likelihood}
   p(d|z) = \prod_{j \in \mathcal{W}_d} p(w_j|f(z)) 
\end{align}
that is, a simple product of word probabilities where the product iterates over all unique words in document $d$ (denoted $\mathcal{W}_d$). In this setting $f(z)$ is a function that maps the hash code, $z$, to a latent vector useful for modelling word probabilities.

\subsubsection{Variational loss}\label{ss:variational-loss}
The first objective of our model is to maximize the document log likelihood:
\begin{align} 
    \log p(d) = \log \int_{\{0,1\}^m} p(d|z)p(z) dz
\end{align}
However, due to the non-linearity of computing $p(w_j|f(z))$ from Equation \ref{eq:doc_likelihood} this computation is intractable and the variational lower bound \cite{kingma2013auto} is maximized instead:
\begin{align}\label{trick1}
    \log p(d) \geq& E_{Q}[\log p(d|z)] - \textrm{KL}( Q(z|d) || p(z))
\end{align}
where $Q(z|d)$ is a learned approximation of the posterior distribution $p(z|d)$, the computation of which we describe in Section \ref{ss:encoder-function}, and $\textrm{KL}$ is the Kullback-Leibler divergence. Writing this out using the document likelihood we obtain the model's variational loss:
\begin{align}
    \mathcal{L}_{\textrm{var}} = E_{Q}\big[\sum_{j \in \mathcal{W}_d} \log  p(w_j|f(z))\big] - \textrm{KL}( Q(z|d) || p(z)) \label{eq:l_var}
\end{align}
where $j$ iterates over all unique words in document $d$. The purpose of this loss is to maximize the document likelihood under our modelling assumptions, where the $E_Q$ term can be considered the reconstruction loss. The KL divergence acts as a regularizer by penalizing large differences between the approximate posterior distribution and the Bernoulli distribution with equal probability of sampling 0 and 1 ($p=0.5$), which can be computed in closed form as:
\begin{align}
    \textrm{KL}( Q(z|d) || p(z)) = Q(d) \log \frac{Q(d)}{p} + (1 - Q(d)) \log \frac{1 - Q(d)}{1 - p}
\end{align}

\subsection{Encoder function}
\label{ss:encoder-function}
The approximate posterior distribution $Q(z|d)$ can be considered as the encoder function that transforms the original document representation into its hash code of $m$ bits. We model this using a neural network that outputs the sampling probabilities used for the Bernoulli sampling of the hash code. First, we compute the representation used as input for computing the sampling probabilities:
\begin{align}
    v_1 &= \textrm{ReLU}(W_a(d \odot E_{\textrm{imp}}) + b_a) \\
    v_2 &= \textrm{ReLU}(W_b v_1 + b_b) 
\end{align}
where $\odot$ corresponds to elementwise multiplication, $W$ and $b$ are weight matrices and bias vectors respectively, and $E_{\textrm{imp}}$ is an \emph{importance embedding} that learns a scalar for each word that is used to scale the word level values of the original document representation, and the same embedding is also used in the decoder function. The purpose of this embedding is to scale the original input such that unimportant words have less influence on the hash code generation. We transform the intermediate $v_2$ representation to a vector of the same size as the hash code, such that the i$^{th}$ entry corresponds to the sampling probability for the i$^{th}$ bit:
\begin{align}
    Q(d) &= \sigma ( W_m v_2 + b_m )
\end{align}
where $W_m$ and $b_m$ have the dimensions corresponding to the code length $m$, and $\sigma$ is the sigmoid function used to enforce the values to be within the interval $[0,1]$, i.e., the range of probability values. The final hash code can then be sampled from the Bernoulli distribution. In practice, this is estimated by a vector $\mu = [\mu_1, \mu_2, ..., \mu_m]$ of values sampled uniformly at random from the interval $[0,1]$ and computing each bit value of either 0 or 1 as:
\begin{align}
    z_i = \lceil Q(d)_i - \mu_i \rceil 
\end{align}
Sampling $\mu$ uniformly at random corresponds to a stochastic strategy, as the same $Q(d)$ could result in different hash codes. The opposite deterministic strategy consists of fixing $\mu_i=0.5$, such that the network always generates the same code for a given document. To encourage exploration during training, the stochastic strategy is chosen, while the deterministic is used for testing. To compute the gradient of the sampled $z$ for back-propagation, we use a straight-through estimator \cite{bengio2013estimating}.

\subsection{Decoder function}
\label{ss:decoder-function}
The purpose of the decoder function is to reconstruct the original document $d$ given the hash code $z$. This is computed using the document log likelihood (Equation \ref{eq:doc_likelihood}) as the sum of word log probabilities:
\begin{align}\label{eq:decoder_function}
    \log p(d|z) &= \sum_{j \in \mathcal{W}_d} \log p(w_j|f(z)) \nonumber \\
    &= \sum_{j \in \mathcal{W}_d} \log \frac{e^{f(z)^T g(E_{\textrm{word}} (o_j \odot E_{\textrm{imp}})) + b_j}}{e^{\sum_{i \in \mathcal{W}_{\textrm{all}}} f(z)^T g(E_{\textrm{word}} (o_i \odot E_{\textrm{imp}})) + b_i}}
\end{align}
where the sums iterate over all unique words in document $d$; $\odot$ corresponds to elementwise multiplication; $o_j$ is a one-hot-vector with 1 in the j$^{th}$ position and 0 everywhere else; $E_{\textrm{imp}}$ is the same importance embedding as in the encoder function; $E_{\textrm{word}}$ is a word embedding; $b$ is a bias vector; $\mathcal{W}_{\textrm{all}}$ contains all vocabulary words; and the $g$ function will be detailed later.
$E_{\textrm{word}}$ is a mapping from a word to a \emph{word embedding} space, such that $\log p(d|z)$ is maximized when the hash code is similar to most words in the document. To this end, the importance embedding assists in reducing the need to be similar to all words, as it learns to reduce the value of unimportant words. The word embedding $E_{\textrm{word}}$ is made by learning a 300 dimensional embedding matrix, and $g(E_{\textrm{word}}(o_j \odot E_{\textrm{imp}}))$ corresponds to a transformation through a fully connected linear layer to fit the code length. The choice of 300 dimensions was made to be similar in size to standard GloVe and Word2vec word embeddings \cite{pennington2014glove,mikolov2013distributed}.
This two-step embedding process was chosen to allow the model to learn a code length-independent embedding initially, such that the underlying word representation is not limited by the code length.

\subsubsection{Reduce overfitting through noise injection}
We inject noise into the hash code before decoding, which has been shown to reduce overfitting and to improve generalizability in generative models \cite{kingma2013auto,sohn2015learning,brock2018large}. For semantic hashing applications, this corresponds to observing significantly more \emph{artificial} documents with small perturbations, which is beneficial for reducing overfitting in the reconstruction step. To this end we choose a Gaussian noise model, which is traditionally done for variational autoencoders \cite{kingma2013auto}, such that $f(z)$ in Equation \ref{eq:decoder_function} is sampled as
$
    f(z) \sim \mathcal{N}(z, \sigma^2 I)
$
where $I$ is the identity matrix and $\sigma^2$ is the variance. Instead of using a fixed variance, we employ variance annealing, where the variance is reduced over time towards 0. Variance annealling has previously been shown to improve performance for generative models in the image domain \cite{brock2018large}, as it reduces the uncertainty over time when the model confidence increases.
However, the gradient estimate with this noise computation exhibits high variance \cite{kingma2013auto}, so we use the reparameterization trick to compute $f(z)$ as:
\begin{align}
    f(z; \sigma^2) = z + \epsilon \sigma^2, \;\;\; \epsilon \sim \mathcal{N}(0, I)
\end{align}
which is based on a single source of normal distributed noise and results in a gradient estimate with lower variance \cite{kingma2013auto}.

\subsection{Ranking component} \label{ss:weak-supervision-modelling}
The variational loss guides the model towards being able to reconstruct the original document from the hash code, but no hash code similarity is enforced between similar documents. We introduce a ranking component into the model, which ensures that similar documents have a small hash code distance between them. To enable the network to learn the correct document ranking we consider document triplets as inputs, ($d$, $d_1$, $d_2$) with corresponding pairwise similarities of $s_{\{d,d_1\}}$ and $s_{\{d,d_2\}}$. However, in the unsupervised setting we do not have a ground truth annotated ranking of the documents to extract the similarities. To this end, we generate pseudo pairwise similarities between the documents, such that weak supervision can be used to train the network in a supervised fashion.

\subsubsection{Estimating pairwise similarities}
For estimating pairwise similarities in our setting, one of many traditional ranking functions or document similarity functions could be employed. We assume such a function is chosen such that a similarity between $d$ and $d_1$ can be computed.

For concreteness, in this paper we choose to compute document similarities using the hash codes generated by Self-Taught Hashing (STH) \cite{zhang2010self} as this has been shown to perform well for semantic hashing (see Section \ref{ss:results}). Using the STH hash codes, document similarity is computed based on the Euclidean distance between two hash codes:
\begin{align}
    s_{\{d,d_1\}} = -|| z^{\textrm{STH}} - z^{\textrm{STH}}_1 ||_2
\end{align}
where $z^{\textrm{STH}}$ corresponds to the STH hash code for document $d$, such that $s_{\{d,d_1\}}$ is highest when two documents are very similar. We use the $k$-nearest neighbour algorithm to find the top $k$ most similar documents for each document.

\subsubsection{Ranking loss}
To train the ranking component we use a modified version of the hinge loss, as the hinge loss has previously been shown to work well for ranking with weak supervision \cite{dehghani2017neural}. We first define the following short-hand expressions:
\begin{align}
    \textrm{sign}_{d,d_1,d_2} &= \textrm{sign}(s_{\{d,d_1\}} - s_{\{d,d_2\}}) \\
    D_{d,d_1,d_2} &= || z - z_2 ||_2^2 - || z - z_1 ||_2^2 
\end{align}
such that $\textrm{sign}_{d,d_1,d_2}$ corresponds to the sign of the estimated pairwise document similarities, and $D_{d,d_1,d_2}$ is the difference between the squared Euclidean distance of the hash codes of the document pairs. Using this we can define our modified hinge loss as the following piece-wise function:
\begin{align}\label{eq:rank-loss}
    \mathcal{L}_{\textrm{rank}} =\begin{cases}
       \max \big( 0, \epsilon- \textrm{sign}_{d,d_1,d_2} D_{d,d_1,d_2} \big)
       &\quad s_{\{d,d_1\}} \neq s_{\{d,d_2\}}\\
       |D_{d,d_1,d_2}|
       &\quad\text{otherwise.} \\ 
     \end{cases}
\end{align}
where $\epsilon$ determines the margin of the hinge loss, which we fix to 1 to allow a small bitwise difference between hash codes of highly similar documents. Traditionally, the hinge loss consists only of the first part of the piece-wise function, but since the similarity estimates are based on distance computations on hash codes, some document pairs will have the same similarity. In that case the pairwise similarities are equal and the loss is simply the absolute value of $D_{d,d_1,d_2}$, as it should be close to 0.

\subsection{Combining variational and ranking components}
\label{ss:combined-components}
We train the variational and ranking components simultaneously by minimizing a combined weighted loss from Equation \ref{eq:l_var} and \ref{eq:rank-loss}:
\begin{align}
    \mathcal{L} &= \alpha \mathcal{L}_{\textrm{rank}} - E_{Q}\big[\sum_{j \in \mathcal{W}_d} \log  p(w_j|f(z))\big] + \beta\textrm{KL}( Q(z|d) || p(z))
\end{align}
where $j$ iterates over all unique words in document $d$, $\alpha$ is used to scale the ranking loss, $\beta$ is used to scale the KL divergence of the variational loss, and we keep the unscaled version of the reconstruction part of the variational loss. During training we start with initial weight parameters of 0 and gradually increase the values in order to focus on just being able to reconstruct the input well.

\section{Experimental evaluation}   

\subsection{Datasets}\label{ss:ddatasets}
We use the four publicly available datasets summarized in Table \ref{tab:datasets}.
1) \emph{20 newsgroups}\footnote{\url{http://scikit-learn.org/0.19/datasets/twenty_newsgroups.html}} is a dataset of posts from 20 different newsgroups. 
2) \emph{TMC}\footnote{\url{https://catalog.data.gov/dataset/siam-2007-text-mining-competition-dataset}} is a dataset of NASA air trafic reports, where each report is labelled with multiple classes.
3) \emph{Reuters21578}\footnote{\url{http://www.nltk.org/book/ch02.html}} is a dataset of news documents from Reuters, where each document is labelled with one or more classes. The Reuters21578 dataset is subsampled such that documents are removed if none of their associated classes are among the 20 most frequent classes. This was done by Chaidaroon and Fang \cite{chaidaroon2017variational} and their subsampled dataset was used by Shen et al. \cite{nash2018}. 4) \emph{AGnews} \cite{zhang2015character} contains news articles from 4 categories.

The datasets are commonly used in related work \cite{chaidaroon2017variational,chaidaroon2018deep,nash2018}, but without full details of preprocessing. So, in the following we describe how we preprocess the data. We filter all documents in a dataset by removing \textit{hapax legomena}, as well as words occurring in more than 90\% of the documents. In addition, we apply stopword removal using the NLTK stopword list\footnote{\url{https://www.nltk.org/nltk_data/}}, do not apply any stemming, and use TF-IDF \cite{salton1988term} as the document representation.

For each dataset we make a training, validation, and testing split of 80\%, 10\%, and 10\% of the data, respectively. In all experiments the training data is used to train an unsupervised model, the validation data is used for early stopping by monitoring when the validation loss starts increasing, and the results are reported on the testing data.

\begin{table}[]
    \centering
    \scalebox{0.95}{
    \begin{tabular}{lcccc}
        \toprule
         & $n$ & multi-class & num. classes & unique words \\ \hline
         20news & 18,846 & No & 20 & 52,447 \\
         TMC & 28,596 & Yes &  22 & 18,196 \\
         Reuters & 9,848 & Yes & 90 & 16,631 \\
         AGnews & 127,598 & No & 4 & 32,154 \\
        \bottomrule
    \end{tabular}
    }
    \caption{Dataset statistics}
    \label{tab:datasets}
    \vspace{-15pt}
\end{table}

\subsection{Performance metric}
The purpose of generating binary hash codes (of equal length) is to use them to obtain fast similarity searches via the Hamming distance, i.e., computing the number of bits where they differ. If two documents are semantically similar, then the generated semantic hash codes should have small Hamming distance between them.
To evaluate the effectiveness of a semantic hashing method we treat each testing document as a query and perform a k-nearest-neighbour (kNN) search using the Hamming distance on the hash codes. Similarly to previous work \cite{chaidaroon2017variational,chaidaroon2018deep,nash2018}, we retrieve the 100 most similar documents and measure the performance on a specific test document as the precision among the 100 retrieved documents (Prec@100). The total performance for a semantic hashing method is then simply the average Prec@100 across all test documents.
The used datasets are originally created for text classification, but we can define two documents to be similar if they share at least one class in their labelling, meaning that multiclass documents need not to be of exactly the same classes. This definition of similarity is also used by related work \cite{chaidaroon2017variational,chaidaroon2018deep,nash2018}.

\subsection{Baselines}
We compare our method against traditional baselines and state-of-the-art semantic hashing methods used in related work as described in Section \ref{s:related-work}: Spectral Hashing (SpH) \cite{weiss2009spectral}, Self-Taught Hashing (STH) \cite{zhang2010self}, Laplacian co-hashing (LCH) \cite{zhang2010laplacian}, Variational Deep Semantic Hashing (VDSH) \cite{chaidaroon2017variational}, NASH \cite{nash2018}, and the neighbourhood recognition model (NbrReg) proposed by Chaidaroon et al. \cite{chaidaroon2018deep}.
We tune the hyperparameters of these methods on the validation data as described in their original papers. 

\subsection{Tuning}
For the encoder function (Section \ref{ss:encoder-function}) we use two fully connected layers with 1000 nodes in each layer on all datasets. The network is trained using the ADAM optimizer \cite{kingma2014adam}. We tune the learning rate from the set $\{0.001, 0.0005\}$, where 0.0005 was chosen consistently for 20news and 0.001 on the other datasets. 
To improve generalization we add Gaussian distributed noise to the hash code before reconstruction in the decoder function, where the variance of the sampled noise distribution is annealed over time. Initially we start with a variance of 1 and reduce it by $10^{-6}$ every iteration, which we choose conservatively to not reduce it too fast. 
For the ranking component we use STH \cite{zhang2010self} to obtain a ranking of the most similar documents for each document in the training and validation set, where we choose every 10$^{th}$ document from the top 200 most similar documents. This choice was made to limit the number of triplets generated for each document as it scales quadraticly in the number of similar documents to consider.

When combining the variational and ranking components of our model (Section \ref{ss:combined-components}), we added a weight parameter on the ranking loss and the KL divergence of the variational loss. We employ a strategy similar to variance annealing in this setting, however in these cases we start at an initial value and increase the weight parameters with very iteration. For the KL divergence we fix the start value at 0 and increase it by $10^{-5}$ with every iteration. For the ranking loss we tune the models by considering starting values from the set $\{0, 0.5, 1, 1.5\}$ and increase from the set $\{30000^{-1}, 300000^{-1}, 1500000^{-1}, 3000000^{-1}\}$.
The code was implemented using the Tensorflow Python library \cite{abadi2016tensorflow} and the experiments were performed on Titan X GPUs.

\subsection{Results}\label{ss:results}
The experimental comparison between the methods is summarized in Table \ref{tab:comparison}, where the methods are used to generate hash codes of length $m \in \{8, 16, 32, 64, 128\}$. We highlight the best performing method according to the Prec@100 metric on the testing data. We perform a paired two tailed t-test at the 0.05 level to test for statistical significance on the Prec@100 scores from each test document. We apply a Shapiro-Wilk test at the 0.05 level to test for normality, which is passed for all methods across all code lengths.

\subsubsection{Baseline comparison}
On all datasets and across all code lengths (number of bits) our proposed Ranking based Semantic Hashing (RBSH) method outperforms both traditional approaches (SpH, STH, and LCH) and more recent neural models (VDSH, NbrReg, and NASH). Generally, we observe a larger performance variation for the traditional methods depending on the dataset compared to the neural approaches, which are more consistent in their relative performance across the datasets. For example, STH is among the top performing methods on Agnews, but performs among the worst on 20news. This highlights a possible strength of neural approaches for the task of semantic hashing. 

Our RBSH consistently outperforms other methods to such a degree, that it generally allows to use hash codes with a factor of 2-4x fewer bits compared to state-of-the-art methods, while keeping the same performance.
This provides a notable benefit on large-scale similarity searches, as computing the Hamming distance between two hash codes scales linearly with the code length. Thus, compared to prior work our RBSH enables both a large speed-up as well as a large storage reduction. 

\subsubsection{Performance versus hash code length}
We next consider how performance scales with the hash code length. For all methods 128 bit codes perform better than 8 bit codes, but the performance of scaling from 8 to 128 bits varies. The performance of SpH and STH on Reuters peaks at 32 bit and reduces thereafter, and a similar trend is observed for VDSH on Agnews and TMC. This phenomenon has been observed in prior work \cite{chaidaroon2017variational,nash2018}, and we posit that it is due to longer hash codes being able to more uniquely encode each document, thus resulting in a degree of overfitting.
However, generally a longer hash code leads to better performance until the performance flattens after a certain code length, which for most methods happens at 32-64 bits.

\subsubsection{Result differences compared to previous work}
Comparing our experimental results to results reported in previous work \cite{chaidaroon2017variational,chaidaroon2018deep,nash2018}, we observe some smaller differences most likely due to preprocessing. Previous work have not fully described the preprocessing steps used, thus to do a complete comparison we had to redo the preprocessing as detailed in Section \ref{ss:ddatasets}.

On 20news and TMC the baseline performance scores we report in this paper are slightly larger for most hash code lengths. The vectorized (i.e., bag-of-words format) Reuters dataset released by the VDSH authors\footnote{\url{https://github.com/unsuthee/VariationalDeepSemanticHashing/blob/master/dataset/reuters.tfidf.mat}}, and also used in the NASH \cite{nash2018} paper, only consisted of 20 (unnamed) classes instead of the reported 90 classes, so these results are not directly comparable.

\begin{table*}
    \centering
    \scalebox{1}{
    \begin{tabular}{@{}llllll|lllll@{}}
    \toprule
      & \multicolumn{5}{c}{20news} & \multicolumn{5}{c}{Agnews}\\ 
         & 8 bits & 16 bits & 32 bits & 64 bits & 128 bits 
         & 8 bits & 16 bits & 32 bits & 64 bits & 128 bits \\ \hline
         
SpH \cite{weiss2009spectral} & 0.0820 & 0.1319 & 0.1696 & 0.2140 & 0.2435 & 0.3596 & 0.5127 & 0.5447 & 0.5265 & 0.5566\\
STH  \cite{zhang2010self} & 0.2695 & 0.4112 & 0.5001 & 0.5193 & 0.5119 & 0.6573 & 0.7909 & 0.8243 & 0.8377 & 0.8378\\
LCH \cite{zhang2010laplacian} & 0.1286 & 0.2268 & 0.4462 & 0.5752 & 0.6507 & 0.7353 & 0.7584 & 0.7654 & 0.7800 & 0.7879\\
VDSH \cite{chaidaroon2017variational} & 0.3066 & 0.3746 & 0.4299 & 0.4403 & 0.4388 & 0.6418 & 0.6754 & 0.6845 & 0.6802 & 0.6714\\
NbrReg \cite{chaidaroon2018deep} & 0.4267 & 0.5071 & 0.5517 & 0.5827 & 0.5857 & 0.4274 & 0.7213 & 0.7832 & 0.7988 & 0.7976\\
NASH  \cite{nash2018} & 0.3537 & 0.4609 & 0.5441 & 0.5913 & 0.6404 & 0.7207 & 0.7839 & 0.8049 & 0.8089 & 0.8142\\ \hline
RBSH  & \textbf{0.5190}$^{\blacktriangle}$ & \textbf{0.6087}$^{\blacktriangle}$ & \textbf{0.6385}$^{\blacktriangle}$ & \textbf{0.6655}$^{\blacktriangle}$ & \textbf{0.6668}$^{\blacktriangle}$ & \textbf{0.8066}$^{\blacktriangle}$ & \textbf{0.8288}$^{\blacktriangle}$ & \textbf{0.8363}$^{\blacktriangle}$ & \textbf{0.8393}$^{\blacktriangle}$ & \textbf{0.8381}$^{\blacktriangle}$\\

    \bottomrule
      & \multicolumn{5}{c}{Reuters} & \multicolumn{5}{c}{TMC}\\ 
         & 8 bits & 16 bits & 32 bits & 64 bits & 128 bits 
         & 8 bits & 16 bits & 32 bits & 64 bits & 128 bits \\ \hline
         
SpH \cite{weiss2009spectral} & 0.4647 & 0.5250 & 0.6311 & 0.5985 & 0.5880 & 0.5976 & 0.6405 & 0.6701 & 0.6791 & 0.6842\\
STH  \cite{zhang2010self} & 0.6981 & 0.7555 & 0.8050 & 0.7984 & 0.7748 & 0.6787 & 0.7218 & 0.7695 & 0.7818 & 0.7797\\
LCH \cite{zhang2010laplacian} & 0.5619 & 0.6235 & 0.6587 & 0.6610 & 0.6586 & 0.6546 & 0.7028 & 0.7498 & 0.7817 & 0.7948\\
VDSH \cite{chaidaroon2017variational} & 0.6371 & 0.6686 & 0.7063 & 0.7095 & 0.7129 & 0.6989 & 0.7300 & 0.7416 & 0.7310 & 0.7289\\
NbrReg \cite{chaidaroon2018deep} & 0.5849 & 0.6794 & 0.6290 & 0.7273 & 0.7326 & 0.7000 & 0.7012 & 0.6747 & 0.7088 & 0.7862\\
NASH  \cite{nash2018} & 0.6202 & 0.7068 & 0.7644 & 0.7798 & 0.8041 & 0.6846 & 0.7323 & 0.7652 & 0.7935 & 0.8078\\ \hline
RBSH  & \textbf{0.7409}$^{\blacktriangle}$ & \textbf{0.7740}$^{\blacktriangle}$ & \textbf{0.8149}$^{\blacktriangle}$ & \textbf{0.8120}$^{\blacktriangle}$ & \textbf{0.8088}$^{\blacktriangle}$ & \textbf{0.7620}$^{\blacktriangle}$ & \textbf{0.7959}$^{\blacktriangle}$ & \textbf{0.8138}$^{\blacktriangle}$ & \textbf{0.8224}$^{\blacktriangle}$ & \textbf{0.8193}$^{\blacktriangle}$\\

    \bottomrule
    \end{tabular}}
    \caption{Prec@100 with varying bit size. Bold marks the highest score. $\blacktriangle$ shows statistically significant improvements with respect to the best baseline at the 0.05 level using a paired two tailed t-test. A Shapiro-Wilk test at the 0.05 level is used to test for normality. }
    \label{tab:comparison}
    \vspace{-15pt}
\end{table*}

\begin{table*}
    \centering
    \scalebox{1}{
    \begin{tabular}{@{}llllll|lllll@{}}
    \toprule
      & \multicolumn{5}{c}{20news} & \multicolumn{5}{c}{Agnews}\\ 
         & 8 bits & 16 bits & 32 bits & 64 bits & 128 bits 
         & 8 bits & 16 bits & 32 bits & 64 bits & 128 bits \\ \hline

RBSH  & \underline{\textbf{0.5190}} & \underline{\textbf{0.6087}} & \underline{\textbf{0.6385}} & \underline{\textbf{0.6655}} & \underline{\textbf{0.6668}} & \underline{\textbf{0.8066}} & \underline{\textbf{0.8288}} & \underline{\textbf{0.8363}} & \underline{\textbf{0.8393}} & \underline{\textbf{0.8381}}\\
RBSH w/o ranking  & \underline{0.4482} & 0.5000 & \underline{0.6263} & \underline{0.6641} & \underline{0.6659} & \underline{0.7986} & \underline{0.8244} & \underline{0.8344} & 0.8332 & 0.8306\\

    \bottomrule
      & \multicolumn{5}{c}{Reuters} & \multicolumn{5}{c}{TMC}\\ 
         & 8 bits & 16 bits & 32 bits & 64 bits & 128 bits 
         & 8 bits & 16 bits & 32 bits & 64 bits & 128 bits \\ \hline

RBSH  & \underline{\textbf{0.7409}} & \underline{\textbf{0.7740}} & \underline{\textbf{0.8149}} & \underline{\textbf{0.8120}} & \underline{\textbf{0.8088}} & \underline{\textbf{0.7620}} & \underline{\textbf{0.7959}} & \underline{\textbf{0.8138}} & \underline{\textbf{0.8224}} & \underline{\textbf{0.8193}}\\
RBSH w/o ranking  & \underline{0.7061} & \underline{0.7701} & \underline{0.8075} & \underline{0.8099} & \underline{0.8081} & \underline{0.7310} & \underline{0.7804} & \underline{0.8040} & \underline{0.8119} & \underline{0.8172}\\

    \bottomrule
    \end{tabular}}
    \caption{Effect of including the ranking component. Prec@100 with varying bit size. Bold marks the highest score and underline marks a score better than the best baseline.}
    \label{tab:effect-ranking}
    \vspace{-20pt}
\end{table*}

\subsection{Effect of ranking component}\label{ss:effect-ranking-component}
To evaluate the influence of the ranking component in RBSH we perform an experiment where the weighting parameter of the ranking loss was set to 0 (thus removing it from the model), and report the results in Table \ref{tab:effect-ranking}. Generally, we observe that on all datasets across all hash code lengths, RBSH outperforms RBSH without the ranking component. However, it is interesting to consider the ranking component's effect on performance as the hash code length increases. On all datasets we observe the largest improvement on 8 bit hash codes, but then on Reuters, Agnews, and TMC a relatively large performance increase happens that reduces the difference in performance. On 20news the performance difference is even larger at 16 bit than at 8 bit, but as the bit size increases the difference decreases until it is marginal. This highlights that one of the major strengths of RBSH, its performance using short hash codes, can be partly attributed to the ranking component. This is beneficial for the application of similarity search, as the Hamming distance scales linearly with the number of bits in the hash codes, and can thus provide a notable speed-up while obtaining a similar performance using fewer bits. Additionally, when comparing the performance of RBSH without the ranking component against the baselines in Table \ref{tab:comparison}, then it obtains a better performance in 17 out of 20 cases, thus highlighting the performance of just the variational component.

To further investigate the ranking component effect, as well as RBSH in general, in Section \ref{ss:importance-emb} we consider word level differences in the learned importance embeddings, as well as relations between inverse document frequency (IDF) and the importance embedding weights for each word. In Section \ref{ss:word-recon} we investigate what makes a word difficult to reconstruct (i.e., using the decoder function in Section \ref{ss:decoder-function}), which is done by comparing the word level reconstruction log probabilities to both IDF and the learned importance embedding weights. Finally, in Section \ref{ss:hash-code-viz} we do a quantitative comparison of RBSH with and without the ranking component. The comparison is based on a t-SNE \cite{maaten2008visualizing} dimensionality reduction of the hash codes, such that a visual inspection can be performed.
In the following sections we consider 16 and 128 bit hash codes generated on 20news, as these provide the largest and one of the smallest performance difference of RBSH with and without the ranking component, respectively.

\subsection{Investigation of the importance embedding}\label{ss:importance-emb}
\begin{figure}
    \centering
    \includegraphics[width=0.41\linewidth]{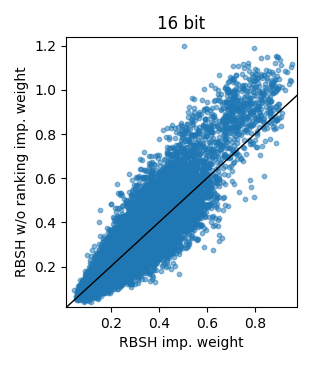}
    \includegraphics[width=0.41\linewidth]{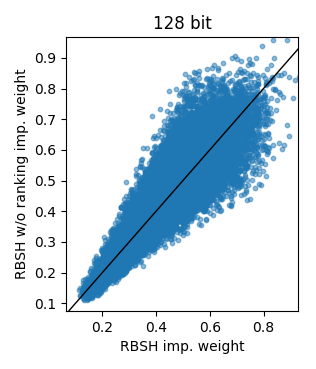}
    \vspace{-12pt}
    \caption{Visualization of the learned importance embedding for each word with and without using the ranking component of RBSH. The plot is made on 20news with 16 and 128 bit hash codes, and the black diagonal line corresponds to equal importance weights.}
    \label{fig:imp_imp}
    \vspace{-15pt}
\end{figure}

We posit that the ranking component in RBSH enables the model to better differentiate between the importance of individual words when reconstructing the original document. If we consider the decoder function in Equation \ref{eq:decoder_function}, then it is maximized when the hash code is similar to most of the importance weighted words, which in the case of equally important words would correspond to a word embedding average. However, if the hash code is short, e.g.,\ 8 bits, then similar documents have a tendency to hash to exactly the same code, as the space of possible codes are considerably smaller than at e.g.,\ 128 bits. 
This leads to worse generalizability observed on unseen documents when using short hash codes, but the ranking component enables the model to better prioritize which words are the most important. Figure \ref{fig:imp_imp} compares the learned importance embedding weights for 16 and 128 bit codes on 20news with and without the ranking component. For 16 bit codes we observe that RBSH without the ranking component tends to estimate a higher importance for most words, and especially for words with an
 RBSH importance over 0.6. This observation could be explained by the ranking component acting as a regularizer, by enabling a direct modelling of which words are important for correctly ranking documents as opposed to just reconstruction. However, as the code length increases this becomes less important as more bits are available to encode more of the occurring words in a document, which is observed from the importance embedding comparison for 128 bits, where the over estimation is only marginal. 

\begin{figure}
    \centering
    \includegraphics[width=0.41\linewidth]{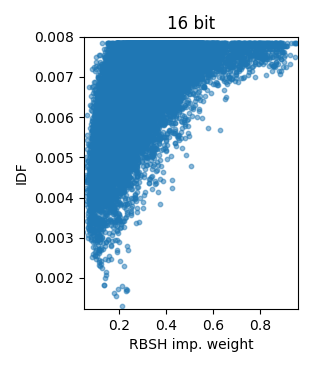}
    \includegraphics[width=0.41\linewidth]{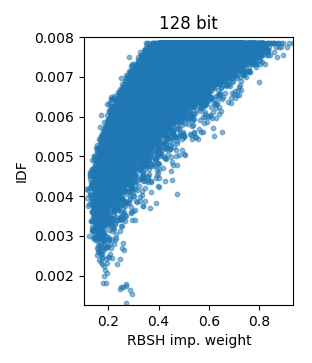}
    \vspace{-12pt}
    \caption{Visualization of the learned importance embedding for each word compared to the inverse document frequency (IDF). The plot is made on 20news with 16 and 128 bit hash codes.}
    \label{fig:imp_idf}
    \vspace{-15pt}
\end{figure}

Figure \ref{fig:imp_idf} shows the importance embedding weights compared to the inverse document frequency (IDF) of each word. For both 16 and 128 bits we observe a similar trend of words with high importance weight that also have a high IDF; however words with a high IDF do not necessarily have a high importance weight. When we consider low importance weights, then the corresponding IDF is more evenly distributed, especially for 16 bit hash codes. For 128 bit we observe that lower importance weights are more often associated with a low IDF. These observations suggest that the model learns to emphasize rare words, as well as words of various rarity that the model deems important for both reconstruction and ranking.


\subsection{Investigation of the difficulty of word reconstruction}\label{ss:word-recon}
\begin{figure}
    \centering
    \includegraphics[width=0.41\linewidth]{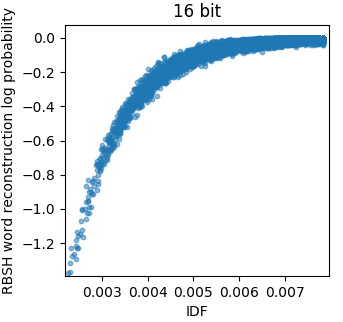}
    \includegraphics[width=0.41\linewidth]{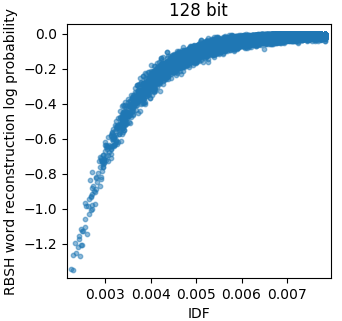}
    \vspace{-12pt}
    \caption{Comparison of the word level reconstruction log probability compared to each word's inverse document frequency (IDF). The plot is made on 20news with 16 and 128 bit hash codes.}
    \label{fig:reconloss}
    \vspace{-15pt}
\end{figure}

To better understand what makes a word difficult to reconstruct we study the word level reconstruction log probabilities, i.e., each summand in Equation \ref{eq:decoder_function}, where a 0 value represents a word that is always possible to reconstruct while a smaller value corresponds to a word more difficult to reconstruct. Figure \ref{fig:reconloss} compares the word level reconstruction log probabilities to each word's IDF for 16 and 128 bit hash codes. There is no notable difference between the plots, which both show that the model prioritizes being able to reconstruct rare words, while focusing less on words occurring often. This follows our intuition of an ideal semantic representation, as words with a high IDF are usually more informative than those with a low IDF.

\begin{figure}
    \centering
    \includegraphics[width=0.41\linewidth]{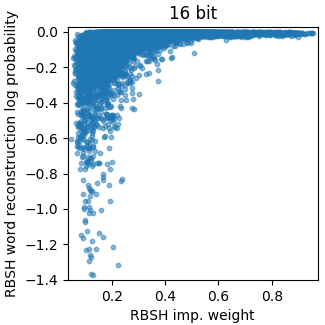}
    \includegraphics[width=0.41\linewidth]{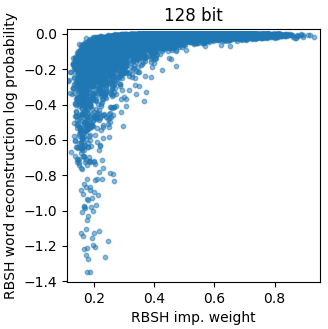}
    \vspace{-12pt}
    \caption{Comparison of the word level reconstruction log probability compared to each word's learned importance weighting. The plot is made on 20news with 16 and 128 bit hash codes.}
    \label{fig:reconloss_imp}
    \vspace{-15pt}
\end{figure}

Figure \ref{fig:reconloss_imp} shows a comparison similar to above, where the word level reconstruction log probabilities are plotted against the learned importance embedding weights. For both 16 and 128 bit hash codes we observe that words that are difficult to reconstruct (i.e., have a low log probability) are associated with a low importance weight. Words with a low reconstruction log probability are also associated with a low IDF. This shows that the model chooses to ignore often occurring words with low importance weight. When considering words with a reconstruction log probability close to 0, then in the case of 16 bit hash codes the corresponding important weights are very evenly distributed in the entire range. In the case of 128 bit hash codes we observe that words the model reconstructs best have importance weights in the slightly higher end of the spectrum, however for lower log probabilities the two hash code lengths behave similarly. This shows that the model is able to reconstruct many words well irrespectively of their learned importance weight, but words with a high importance weight are always able to be reconstructed well. 


\begin{figure*}[]
    \centering
    \includegraphics[width=0.36\linewidth]{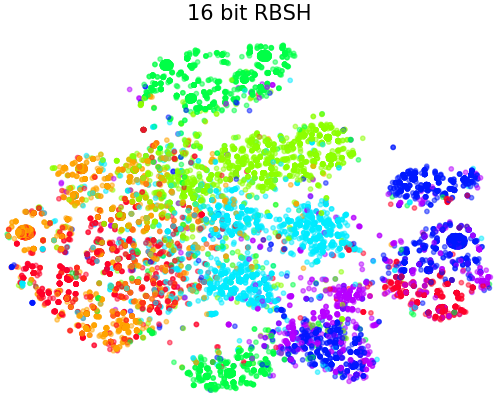}
    \hspace{25pt}
    \includegraphics[width=0.36\linewidth]{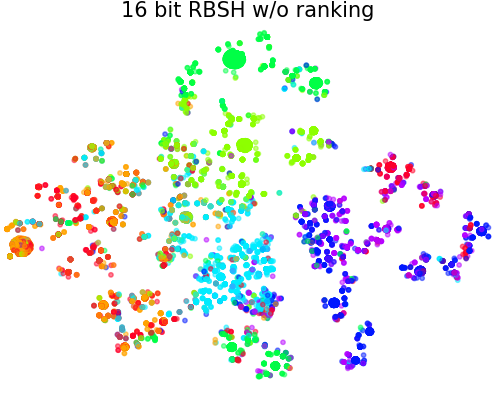}

    \includegraphics[width=0.36\linewidth]{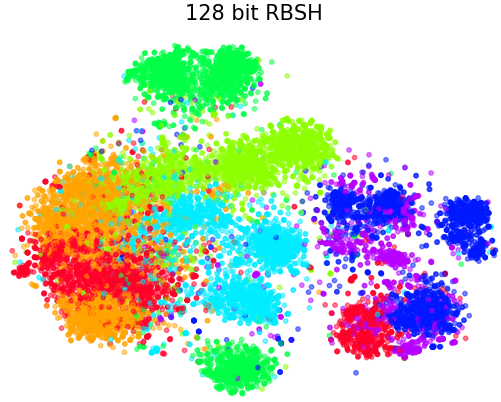}
    \hspace{25pt}
    \includegraphics[width=0.36\linewidth]{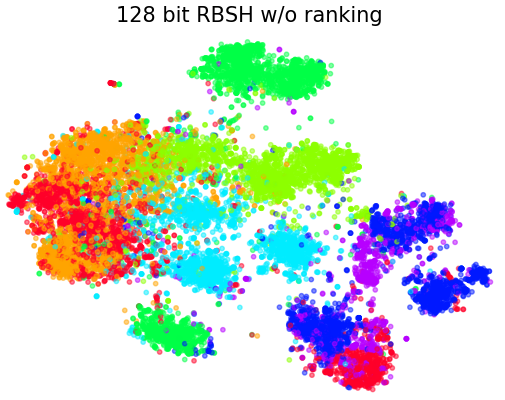}
    \vspace{-10pt}
    \caption{t-SNE \cite{maaten2008visualizing} visualization of the 16 and 128 bit hash codes from our RBSH with and with the ranking component. 20news was used as the dataset and the same color coding for class labels is used across the plots.}
    \label{fig:clustering}
    \vspace{-10pt}
\end{figure*}

\subsection{Hash code visualization}\label{ss:hash-code-viz}
In Section \ref{ss:importance-emb} we argued that the ranking component of RBSH enables the model to better prioritize important words for short hash codes, by directly modelling which words were relevant for ranking the documents. To further study this we perform a qualitative visualization using t-SNE \cite{maaten2008visualizing} of 16 and 128 bit hash codes on 20news (see Figure \ref{fig:clustering}), where we do the visualization for RBSH with and without the ranking component. For 16 bit hash codes we observe that RBSH without the ranking component most often creates very tight clusters of documents, corresponding to the fact that many of the produced hash codes are identical. When the ranking component is included the produced hash codes are more varied. This creates larger, more general clusters of similar documents. This leads to better generalizability as the space is better utilized, such that unseen documents are less likely to hash into unknown regions, which would result in poor retrieval performance. When considering the 128 bit hash codes for RBSH with and without the ranking component, we observe that they are highly similar, which was also expected as the Prec@100 performance was almost identical. 


\section{Conclusion}
We presented a novel method for unsupervised semantic hashing, \textit{Ranking based Semantic Hashing} (RBSH), which consists of a variational and ranking component. The variational component has similarities with variational autoencoders and learns to encode a input document to a binary hash code, while still being able to reconstruct the original document well. The ranking component is trained on document triplets and learns to correctly rank the documents based on their generated hash codes. To circumvent the need of labelled data, we utilize a weak labeller to estimate the rankings, and then employ weak supervision to train the model in a supervised fashion. These two components enable the model to encode both local and global structure into the hash code. Experimental results on four publicly available datasets showed that RBSH is able to significantly outperform state-of-the-art semantic hashing methods to such a degree, that RBSH hash codes generally perform similarly to other state-of-the-art hash codes, while using 2-4x fewer bits.
This means that RBSH can maintain state-of-the-art performance while allowing a direct storage reduction of a factor 2-4x.
Further analysis showed that the ranking component provided performance increases on all code lengths, but especially improved the performance on hash codes of 8-16 bits. Generally, the model analysis also highlighted RBSH's ability to estimate the importance of rare words for better hash encoding, and that it prioritizes the encoding of rare informative words in its hash code.

Future work includes incorporating multiple weak labellers when generating the hash code ranking, which under certain independence assumptions has been theoretically shown to improve performance of weak supervision \cite{zamani2018theory}. Additionally, it could be interesting to investigate the effect of more expressive encoding functions, such as recurrent or convolutional neural networks, that have been used for image hashing \cite{zhao2015deep,yao2016deep}.

\begin{acks}
Partly funded by Innovationsfonden DK, DABAI (5153-00004A).
\end{acks}

\newpage
\bibliographystyle{ACM-Reference-Format}
\balance

\end{document}